\documentclass[iop]{emulateapj}
\pdfoutput=1
\usepackage{graphicx}
\usepackage{setspace}
\usepackage{natbib}
\usepackage{color}
\usepackage{amsmath,amssymb}
\usepackage{times}
\usepackage{gensymb}

\begin{document}

\title{Isotropic at the Break? 3D Kinematics of Milky Way Halo Stars in the Foreground of M31}
\author{Emily C. Cunningham\altaffilmark{1}, Alis J. Deason\altaffilmark{2,1}, Puragra Guhathakurta\altaffilmark{1}, Constance M. Rockosi\altaffilmark{1}, Roeland P. van der Marel\altaffilmark{3}, Elisa Toloba\altaffilmark{1,4}, Karoline M. Gilbert\altaffilmark{3}, Sangmo Tony Sohn\altaffilmark{5}, and Claire E. Dorman\altaffilmark{1}}  

\altaffiltext{1}{Department of Astronomy and Astrophysics, University
  of California Santa Cruz, Santa Cruz, CA 95064, USA; eccunnin@ucsc.edu}
\altaffiltext{2}{Kavli Institute for Particle Astrophysics and Cosmology \& Physics Department, Stanford University, Stanford, CA 94305, USA}  
\altaffiltext{3}{Space Telescope Science Institute, 3700 San Martin
  Drive, Baltimore, MD 21218, USA}
\altaffiltext{4}{Texas Tech University, Physics Department, Box 41051, Lubbock, TX 79409-1051, USA}
\altaffiltext{5}{Department of Physics and Astronomy, The Johns Hopkins University, Baltimore, MD 21218, USA}
\date{\today}

\begin{abstract}
We present the line-of-sight (LOS) velocities for 13 distant main sequence Milky Way halo stars with published proper motions. The proper motions were measured using long baseline (5--7 years) multi-epoch \textit{HST}/ACS photometry, and the LOS velocities were extracted from deep (5--6 hour integrations) Keck II/DEIMOS spectra. We estimate the parameters of the velocity ellipsoid of the stellar halo using a Markov chain Monte Carlo ensembler sampler method. The velocity second moments in the directions of the Galactic $(l,b,$ LOS) coordinate system are $\langle v^2_l \rangle^{1/2} = 138^{+43}_{-26}$ km s$^{-1}$, $\langle v^2_b \rangle^{1/2} = 88^{+28}_{-17} \ \mbox{km \ s}^{-1}$, and $\langle v^2_{\rm{LOS}} \rangle^{1/2} = 91^{+27}_{-14} \ \mbox{km \ s}^{-1}$. We use these ellipsoid parameters to constrain the velocity anisotropy of the stellar halo. Ours is the first measurement of the anisotropy parameter $\beta$ using 3D kinematics outside of the solar neighborhood. We find $\beta=-0.3^{+0.4}_{-0.9}$, consistent with isotropy and lower than solar neighborhood $\beta$ measurements by 2$\sigma$ ($\beta_{SN} \sim 0.5-0.7$). We identify two stars in our sample that are likely members of the known TriAnd substructure, and excluding these objects from our sample increases our estimate of the anisotropy to $\beta=0.1^{+0.4}_{-1.0}$, which is still lower than solar neighborhood measurements by $1\sigma$. The potential decrease in $\beta$ with Galactocentric radius is inconsistent with theoretical predictions, though consistent with recent observational studies, and may indicate the presence of large, shell-type structure (or structures) at $r \sim 25$ kpc. The methods described in this paper will be applied to a much larger sample of stars with 3D kinematics observed through the ongoing HALO7D\footnote{Halo Assembly in Lambda-CDM: Observations in 7-Dimensions (HALO7D) is a spectroscopic survey of distant, Milky Way halo stars with Keck II/DEIMOS. The 7 dimensions are the 6 dimensions of phase space plus chemical abundances.} program.  

\end{abstract}

\section{Introduction}
\setcounter{footnote}{0}

The Milky Way halo devours hundreds of lower mass dwarf galaxies over its lifetime. The stripped stellar material from this voracious eating habit is splayed out in a vast, diffuse stellar halo. The orbital timescales at these large distances ($\gtrsim$ 10 kpc) are very long, and the halo stars retain a memory of their initial conditions. Thus, by studying the phase space distribution of halo stars, we are privy to a unique window into the past accretion history of our Galaxy.

Global kinematic properties, such as the velocity anisotropy (i.e., the relative pressure between tangential and radial velocity components), can provide important insight into the formation of the stellar halo (see \citealt{Binney2008}). The exact merger and dissipation history of a spheroid can strongly affect its velocity anisotropy profile (e.g., \citealt{Naab2006}; \citealt{Deason2013a}). Local studies, limited to heliocentric distances $D \lesssim 10$ kpc, have measured the full 3D kinematics of halo stars. This has revealed a strongly radially biased velocity anisotropy with $\beta=1-\sigma^2_{\rm tan}/\sigma^2_{\rm rad} \approx 0.5-0.7$ (e.g., \citealt{Smith2009};  \citealt{Bond2010}), in seemingly good agreement with the predictions of simulations (e.g., \citealt{Bullock2005}; \citealt{Cooper2010}).

In \cite{Deason2013b} (hereafter D13), we exploited the long time-baselines and
exquisite photometry of deep, multi-epoch \textit{HST} fields to
measure the \textit{proper motions} (PMs) of main sequence turn-off
(MSTO) stars in the distant Milky Way halo. Our pilot program used
5--7 year baseline \textit{HST}/ACS fields towards M31 to measure PMs
of $N\sim13$ halo stars in the foreground. Our PMs are extremely accurate,
with random errors of $\sim$ 5 km s$^{-1}$.  These 13 halo stars
provided the first direct bound on the tangential velocity moments of the halo
in this extreme radial regime, and 
provide new insights into halo structure. From the PMs measured for 13 Milky Way halo stars at 18 $\lesssim$ {$r$} $\lesssim$ 30~kpc
in our M31 \textit{HST} fields, D13 inferred approximate isotropy between
radial and tangential motions: $\beta$ = 0.0$^{+0.2}_{-0.4}$. This differs by $3\sigma$ from local measures of the
velocity anisotropy, which find strongly radial orbits. This trend of decreasing radial anisotropy with
galactocentric distance conflicts with numerical simulations, which
predict an outward increase in radial anisotropy.

In D13, we had no line-of-sight (LOS) information for these stars:
we relied on the LOS velocities of other halo tracers (blue horizontal branch (BHB) stars, K giants) in different regions
of the sky to form our argument. With spectroscopic information, we
circumvent the need to rely on independent, and perhaps biased,
tracers.
In this paper, we present the LOS velocities for our halo star candidates, and use this 3D kinematic information to estimate the parameters of the velocity ellipsoid and the velocity anisotropy.

The paper is arranged as follows. In Section \ref{sec:data}, we describe the target selection, proper motion measurements, spectroscopic observations and LOS velocity extraction. In Section \ref{sec:param}, we describe our method for estimating the parameters of the velocity ellipsoid. Our results are presented in Section \ref{sec:res}, and discussed in Section \ref{sec:disc}. We summarize our findings in Section \ref{sec:concl}. 

\begin{figure*}[t]
	\centering
	\includegraphics[width=\textwidth]{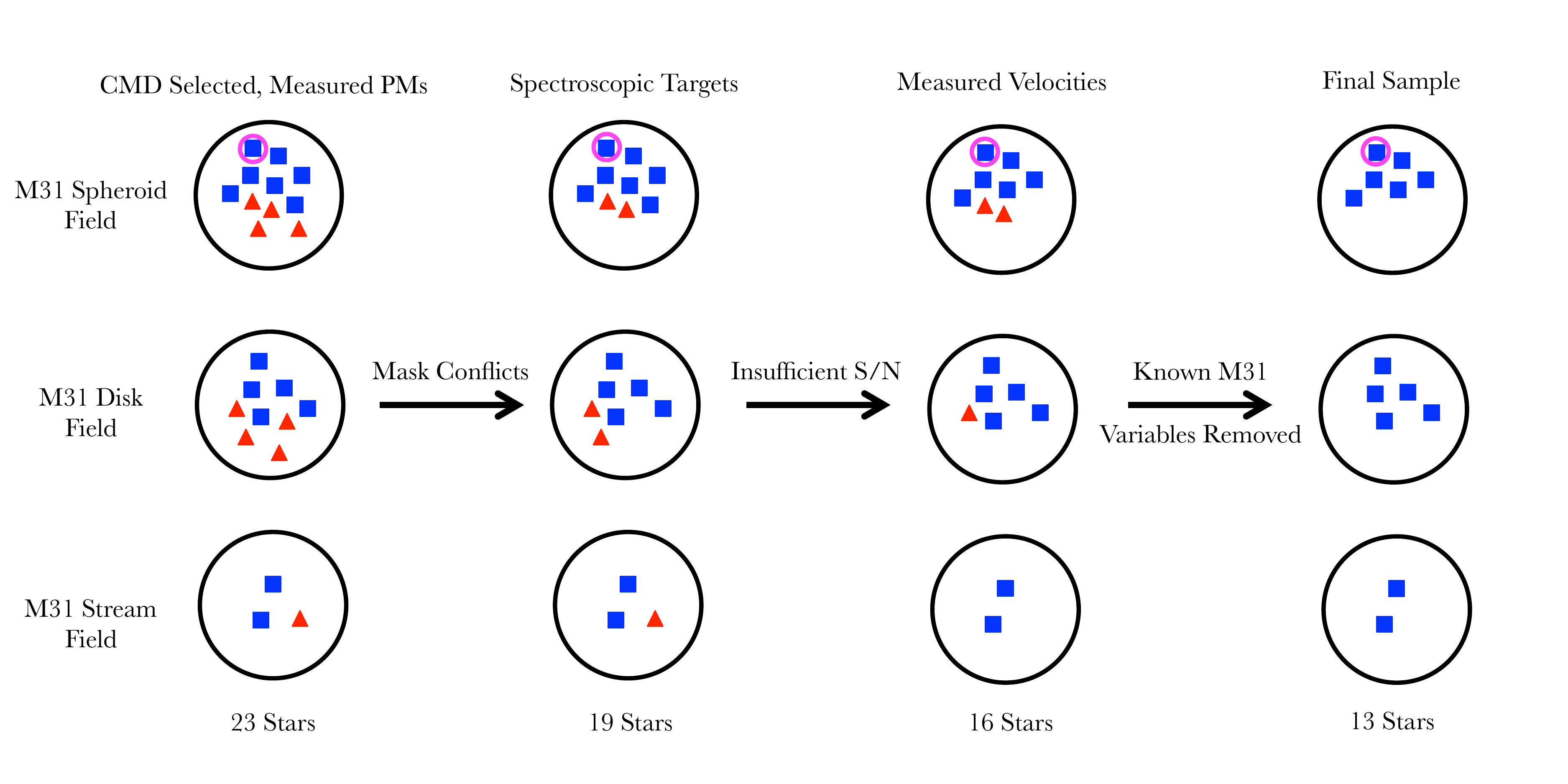}
	\caption[]{Visual representation of the evolution of our sample, from the 23 CMD-selected halo star candidates from D13 to our sample of 13 stars. The different symbols represent the classification of the stars based on their proper motions: red triangles are M31 star candidates, while blue squares are Milky Way stars (see Fig. 3 of D13). The pink circle denotes the object classified as a potential Milky Way disk star in D13.}
	\label{fig:sample}
\end{figure*}

\begin{figure*}[h]
	\centering
	\includegraphics[width=\textwidth]{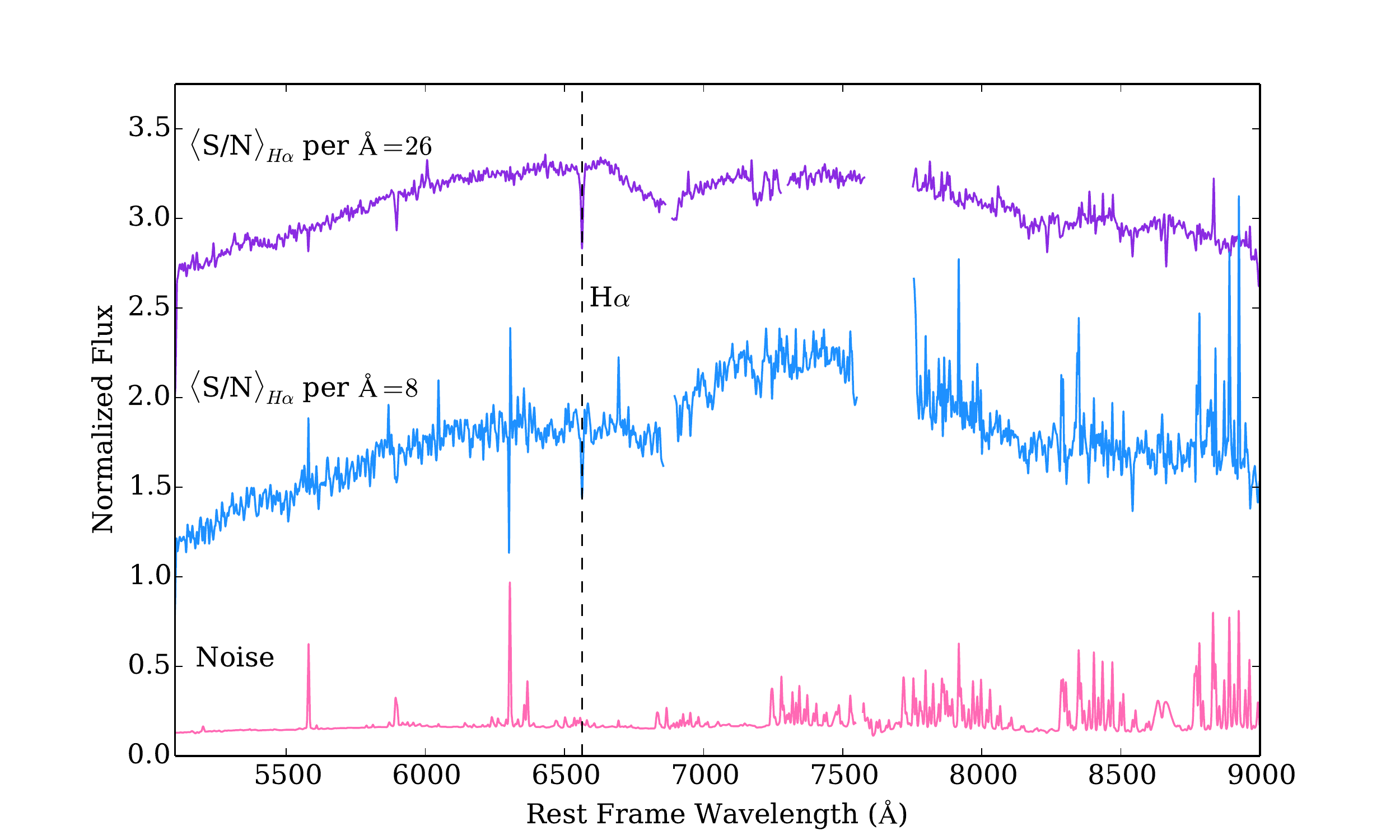}
	\caption[]{Two sample stellar spectra from our sample: one of our higher signal-to-noise spectra (S/N per \AA =26) is shown in purple and a lower signal-to-noise spectrum (S/N per \AA =8) is shown in blue. The noise spectrum for the lower signal-to-noise spectrum is shown on the bottom in pink. The spectra have been normalized, smoothed with a Gaussian kernel with $\sigma=3$, and plotted with a vertical offset. We mask the chip gap in each spectrum, as well as the telluric A band at 7600 $\rm{\AA}$ and the telluric B band at 6875 \AA. The dip in the spectra between 6700 $\rm{\AA}$ and 7200 $\rm{\AA}$ is an instrumental feature. Spikes in the noise spectrum correspond to night sky emission lines. The signal to noise ratios are computed at H$\alpha$. }
	\label{fig:spec}
\end{figure*}

\begin{table*}[h]
\begin{center}
\renewcommand{\tabcolsep}{0.1cm}
\renewcommand{\arraystretch}{0.75}
\begin{tabular}{c  c  c  c  c  c  c c c c}
\hline
Field & RA (J2000) & DEC (J2000) & $m_{\mathrm{F814W}}$ &
  $m_{\mathrm{F606W}}$ & $\mu_l$ [mas yr$^{-1}$] & $\mu_b$ [mas yr$^{-1}$] & $v_{\rm{LOS}}$ (GSR) [km s$^{-1}$] & S/N per \AA \\
\hline
M31 Spheroid & & & &  & & & \\
\hline
& 00:46:01.47  &  +40:41:35.53   &   21.86  &    21.45 & $-1.96 \pm 0.04$&  $-2.08 \pm 0.04$& $54 \pm 8^{*}$ &  19.4 \\ 
& 00:46:03.79  &  +40:41:22.81   &   22.53  &    22.19 & $1.36 \pm 0.02$&  $-1.33 \pm 0.02$& $-68 \pm 17 $ & 9.3 \\ 
& 00:46:03.67  &  +40:41:56.60   &   22.88  &    22.52 & $2.12 \pm 0.03$&  $-0.82 \pm 0.02$& $-90 \pm 19 $ & 9.4 \\
& 00:46:06.41  &  +40:42:15.07   &   22.53  &    22.06 & $1.45 \pm 0.02$& $-0.90 \pm 0.02$ & $ 37 \pm 16 $ & 10.5\\
& 00:46:05.14  &  +40:43:37.19   &   21.82  &    21.47 & $3.91 \pm 0.02$&  $-1.59 \pm 0.02$& $120 \pm 6 $ & 20.9 \\ 
& 00:46:12.92  &  +40:41:22.51   &   22.92  &    22.61 & $1.88 \pm 0.06$&  $-2.83 \pm 0.06$& $43 \pm 15^{*}$ & 8.1 \\
\hline
M31 Disk & & & &  & & & \\
\hline
& 00:49:08.91  &  +42:44:13.62   &   21.79  &    21.40 & $-0.59 \pm 0.03$&   $-1.50 \pm 0.04$& $73 \pm 4 $ & 30.0 \\
& 00:49:08.30  &  +42:44:50.44   &   22.12  &    21.66 & $+1.03 \pm 0.04$&  $-0.78 \pm 0.04$& $-42 \pm 6$ & 26.0 \\
& 00:49:13.50  &  +42:43:36.17   &   22.71  &    22.35 & $-0.71 \pm 0.07$& $-0.67 \pm 0.08$ & $-117 \pm 10$ & 15.9 \\
& 00:49:13.38  &  +42:45:56.93   &   23.62  &    23.30 & $+2.16 \pm 0.05$&  $-0.40 \pm 0.06$& $142 \pm 39$ & 7.2 \\
& 00:49:13.69  &  +42:45:52.07   &   24.76  &    24.29 & $+0.64 \pm 0.07$& $+0.58 \pm 0.06$ & $-175 \pm 10$ & 12.2 \\
\hline
M31 Stream & & & &  & & & \\
\hline
& 00:44:26.44 &   +39:47:33.43   &   22.69  &    22.35 &$+0.00 \pm 0.06$ &$-1.85 \pm 0.06$ & $-89 \pm 7$ & 18.3 \\
& 00:44:23.93 &   +39:46:26.25   &   23.83  &    23.46 &$-0.43 \pm 0.05$ &$-1.13 \pm 0.07$ & $16 \pm 16$ & 6.9 \\
\hline
\end{tabular}
  \caption{\small The properties of the candidate halo stars with measured 3D kinematics used in this analysis. We give the right ascension (RA) and declination (DEC), \textit{HST}/ACS
  STMAG magnitudes, PMs in Galactic coordinates and LOS velocity (in the Galactocentric frame). The RA, DEC and magnitudes come from \cite{Brown2009}, and
  the proper motions derive from the study by \cite{Sohn2012a}. The LOS velocity measurements are described in Section \ref{sec:vel_meas}. Potential TriAnd members are indicated by an asterisk. The signal to noise ratios are computed at H$\alpha$. }
\label{tab:pms}
\end{center}
\end{table*}

\section{Dataset}
\label{sec:data}

\subsection{HST Imaging: Proper Motions}
A detailed description of the target selection can be found in D13, but we summarize the key points here. Our objects were selected from three \textit{HST} observing programs: GO-9453, GO-10265 (PI: T.Brown), and GO-11684 (PI: R.P. van der Marel). The combination of these three programs provide deep, multi-epoch optical imaging of three fields in M31 (M31 Spheroid, M31 Disk and M31 Stream). These observations were used to measure the proper motion of M31 (\citealt{Sohn2012a}), and during the course of this study, proper motion catalogs for \textit{individual} stars in the three \textit{HST} fields were created.

D13 selected Milky Way halo star candidates in color-magnitude space, using photometry from \cite{Brown2009}: all stars fall within $m_{\mathrm{F606W}}-m_{\mathrm{F814W}} \sim
-0.3$ and $21.5 \lesssim m_{\mathrm{F814W}} \lesssim 25.5$. In this region of the color-magnitude diagram (CMD) we expect minimal contamination from the Milky Way disk and M31's red giant branch (see Section 2.2 and Figure 1 of D13). Proper motions were then used to classify the objects as M31, Milky Way halo and Milky Way disk stars. The average uncertainty in the proper motion measurements is $\sigma_\mu \sim 0.05$ mas yr$^{-1}$.

\subsection{Keck/DEIMOS Spectra}
\subsubsection{Spectroscopic Sample}

Figure \ref{fig:sample} demonstrates how our initial sample from D13 evolved into the sample used in this analysis. In D13, we presented proper motions for the 23 candidate halo stars selected from color-magnitude diagrams (CMDs): 11, 9, and 3 stars in the M31 Spheroid, M31 Disk and M31 Stream fields, respectively. Based on the proper motions, 13 of these stars were classified as Milky Way halo stars, 9 as M31 stars, and 1 as a potential Milky Way disk star (see Figure 3 of D13). The symbols in Figure \ref{fig:sample} represent the proper motion classification: Milky Way halo star candidates are blue squares, M31 star candidates are red triangles, and the pink circle denotes the potential Milky Way disk star. We obtained spectra for 19 of the original 23 stars; we were not able to obtain spectra for all of the halo star candidates due to conflicts in the spectral direction on the DEIMOS slitmask. Three additional stars were too faint to measure velocities. After removing known variables in M31 (\citealt{Brown2004}; \citealt{Jeffery2011}), we were left with our final sample of 13 objects. It is worth noting that this is \textit{not} the exact same sample of 13 stars used in the kinematic analysis of D13: one of the objects we used in D13 was very faint ($m_{\rm F814W}=24.05$) and without strong spectral features, so we were unable to measure its velocity. We include the object classified as a potential disk star in D13 in our analysis (as its LOS velocity is consistent with halo kinematics).\footnote{As outlined in Section \ref{sec:vel_meas}, we find that this star is likely a member of TriAnd.} The properties of our 13 stars are summarized in Table \ref{tab:pms}.

\begin{figure*}
\centering
\begin{minipage}{0.47\linewidth}
	\centering
	\includegraphics[width=\textwidth]{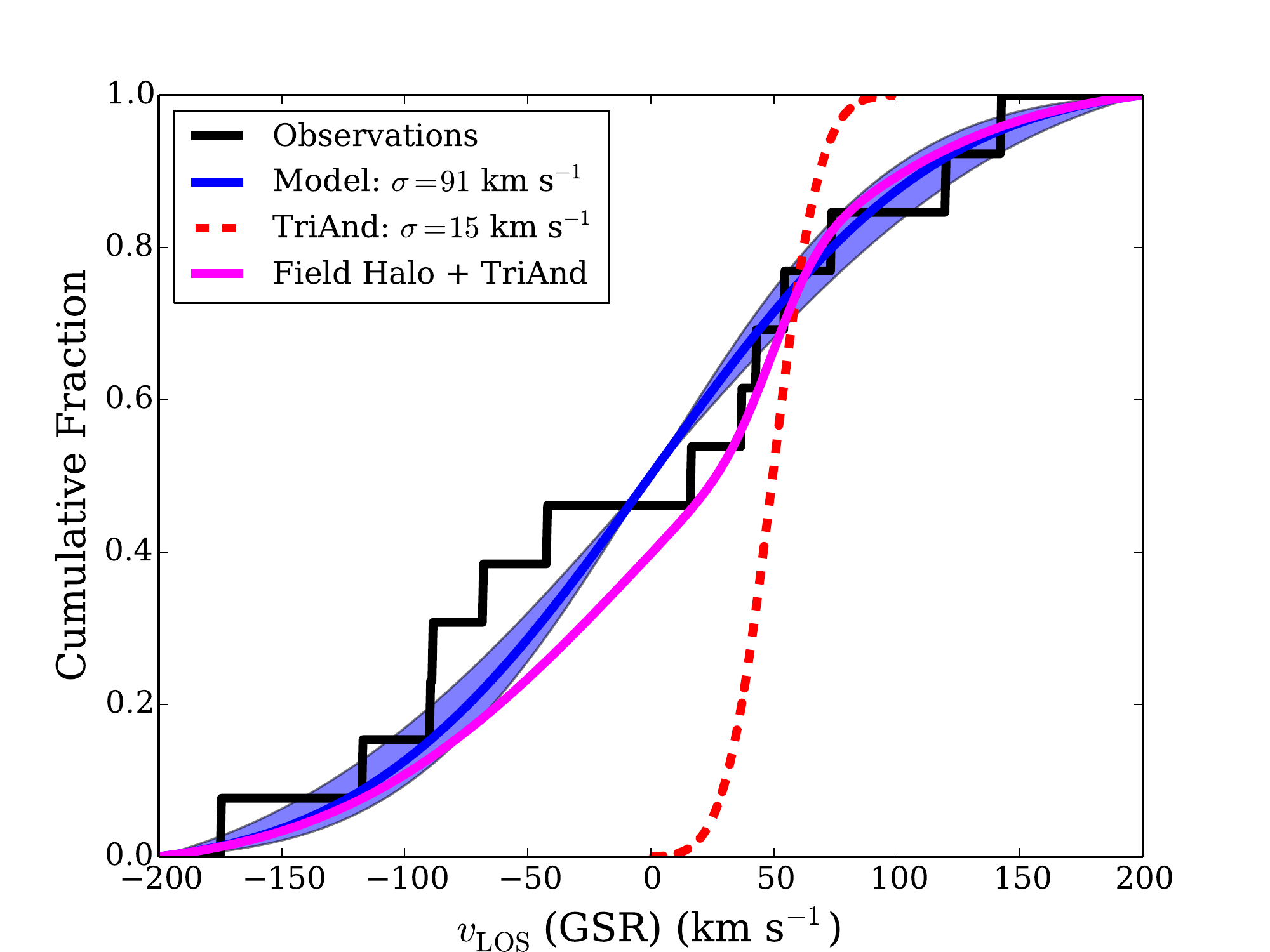}
	\caption{Cumulative histogram of LOS velocities (in the Galactocentric frame) of the 13 halo stars in our sample (black). The overplotted blue line shows the CDF for the most likely value for $\sigma_{\rm{LOS}}$ for the full sample (see Sec. \ref{sec:param}), with the shaded blue region indicating the 68\% confidence region. An approximate CDF for the Triangulum-Andromeda Stream (TriAnd) is shown in red ($v_0 \sim 50$ km s$^{-1}$, $\sigma \sim 15 $ km s$^{-1}$). The pink line shows the CDF when the LOS velocity distribution is modelled as a double Gaussian, with TriAnd ($\sim 20 \%$) and the field halo treated as separate components.}
	\label{fig:vel_hist}
\end{minipage}
\quad
\begin{minipage}{0.47\linewidth}
	\centering
	\includegraphics[width=\textwidth]{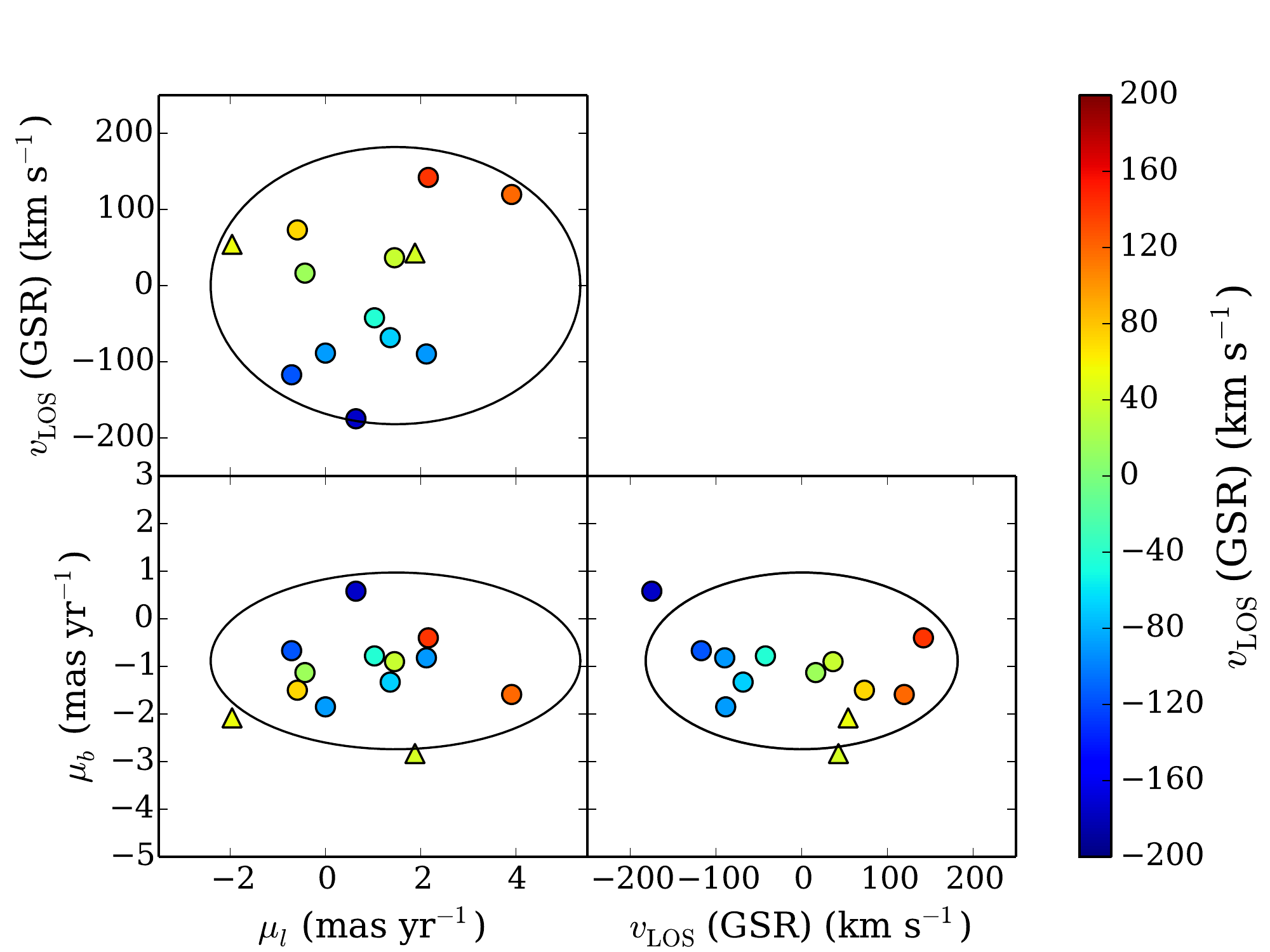}
	\caption{Projections of our 3D kinematic sample, color coded by LOS velocity in the GSR frame. Likely TriAnd members are indicated by triangles. The ellipses show the 2D projection of the 3D velocity ellipsoid; the mean sample distance (20 kpc; see Table \ref{tab:ellipsoid}, Section \ref{sec:param}) was used to convert from km s$^{-1}$ to mas yr$^{-1}$ for the proper motion axes. Ellipses are drawn to enclose 2$\sigma$. }
	\label{fig:mu_mu}
\end{minipage}
\end{figure*}

\subsubsection{Observations}

Observations were taken on September 28--30, 2014 on the Keck II telescope with the DEIMOS spectrograph (\citealt{Faber2003}). Over the course of the run, the seeing varied from $0.45''-0.9''$. We observed one slitmask in each of the three fields with the 600 line/mm grating. The central wavelength was 7200 \AA, resulting in a wavelength range of $\sim4500-9300$ \AA, where the exact wavelength range for each object varies depending on its position on the mask. The spectral resolution at $H \alpha$ (6563 \AA) is $R\sim 2000$ (measured at the FWHM). In order to limit the flux losses due to atmospheric dispersion, we tilted our slits such that the position angle of the slit was consistent with the median parallactic angle of the observing block. The masks in the Spheroid and Disk fields were observed for a total of 5.9 hours, and the Stream field mask was observed for 5.3 hours. The slitmasks were then processed by a modified version of the \textit{spec2d} pipeline developed by the DEEP2 team at UC Berkeley (\citealt{MCooper2012}). Two spectra from our sample are plotted in Figure \ref{fig:spec}; the top spectrum in Figure \ref{fig:spec}, shown in purple, has one of the higher signal-to-noise ratios of our sample (S/N per \AA=26 at H$\alpha$), while the lower spectrum, shown in blue, is an example of one of our lower signal-to-noise objects (S/N per \AA=8 at H$\alpha$). The noise spectrum from the lower signal-to-noise object is shown at the bottom of the figure in pink.

\subsubsection{Velocity Measurements}
\label{sec:vel_meas}

Line-of-sight (LOS) velocities are measured from one-dimensional spectra using the Penalized Pixel-Fitting method (pPXF) of \cite{Cappellari2004}. The program determines the best fit composite stellar template for a given target using a penalized maximum likelihood approach. The 31 stellar templates employed in this analysis are described in detail in Toloba et al. (2016, submitted); the templates have high signal-to-noise ratios (100--800 $\rm{\AA}^{-1}$), and span a range of spectral types (from B1 to M8) and luminosity classes (from dwarfs to supergiants).

Errors in the raw velocity are determined through 1000 Monte Carlo simulations. In each simulation, we perturb the flux of the spectrum by adding noise to each pixel based on the uncertainty of the flux measurement in that pixel. The amount of noise added is drawn from a Gaussian distribution with width equal to the flux uncertainty. We then measure the velocity of each perturbed spectrum, and the error on the LOS velocity is taken to be the biweight standard deviation of the Gaussian distribution of velocities of perturbed spectra.

A-band telluric corrections are measured using the same method, and heliocentric LOS velocities are calculated by applying the A-band and heliocentric corrections to the raw velocities. The final uncertainty in the heliocentric LOS velocity is determined by adding in quadrature the errors on the raw velocity and the A-band correction. 

Figure \ref{fig:vel_hist} shows a cumulative histogram of the LOS velocities for our sample of halo stars, in the frame of the Galactic Standard of Rest (GSR). Observed heliocentric velocities are converted to Galactocentric ones by
assuming a circular speed of 240 km s$^{-1}$ (e.g., \citealt{Reid2009}; \citealt{McMillan2011}; \citealt{Schonrich2012}) at the position of the
sun ($R_0 = 8.5$ kpc) with a solar peculiar motion ($U, V, W$)=(11.1,
12.24, 7.25) km s$^{-1}$ (\citealt{Schonrich2010}). Here, $U$ is directed toward the
Galactic center, $V$ is positive in the direction of Galactic rotation
and $W$ is positive towards the North Galactic Pole. 

In Figure \ref{fig:vel_hist}, we see evidence for a ``hot halo" population: there are no sharp increases where we expect to see contamination from the Milky Way Disk (along this line of sight, $\langle v_{\rm{disk}} \rangle \sim 145$ km s$^{-1}$) or M31 ($\langle v_{\rm{M31}} \rangle \sim -150$ km s$^{-1}$). The blue curve shows the cumulative distribution function (CDF) for the $\sigma_{\rm{LOS}}$ value with maximum posterior probability (see Section \ref{sec:param}), with the shaded blue region indicating the 68\% confidence region. In contrast, as an example of substructure that is dynamically cold in LOS velocity, an approximate CDF for the Triangulum-Andromeda Stream (TriAnd; located along the line-of-sight towards M31) is shown in red ($v_0 \sim 50$ km s$^{-1}$, $\sigma \sim 15 $ km s$^{-1}$; e.g., \citealt{Deason2014b}; \citealt{Sheffield2014}). D13 suggested that the presence of a cold stream or TriAnd could be the reason for the relative increase in tangential pressure seen in this sample. However, our LOS velocity measurements confirm that this is not the case: the significant dispersion in the LOS velocity distribution demonstrates that our sample is not dominated by members of a cold stream nor by TriAnd.

\begin{figure*}[t]
	\centering
	\includegraphics[width=13cm, height=13cm]{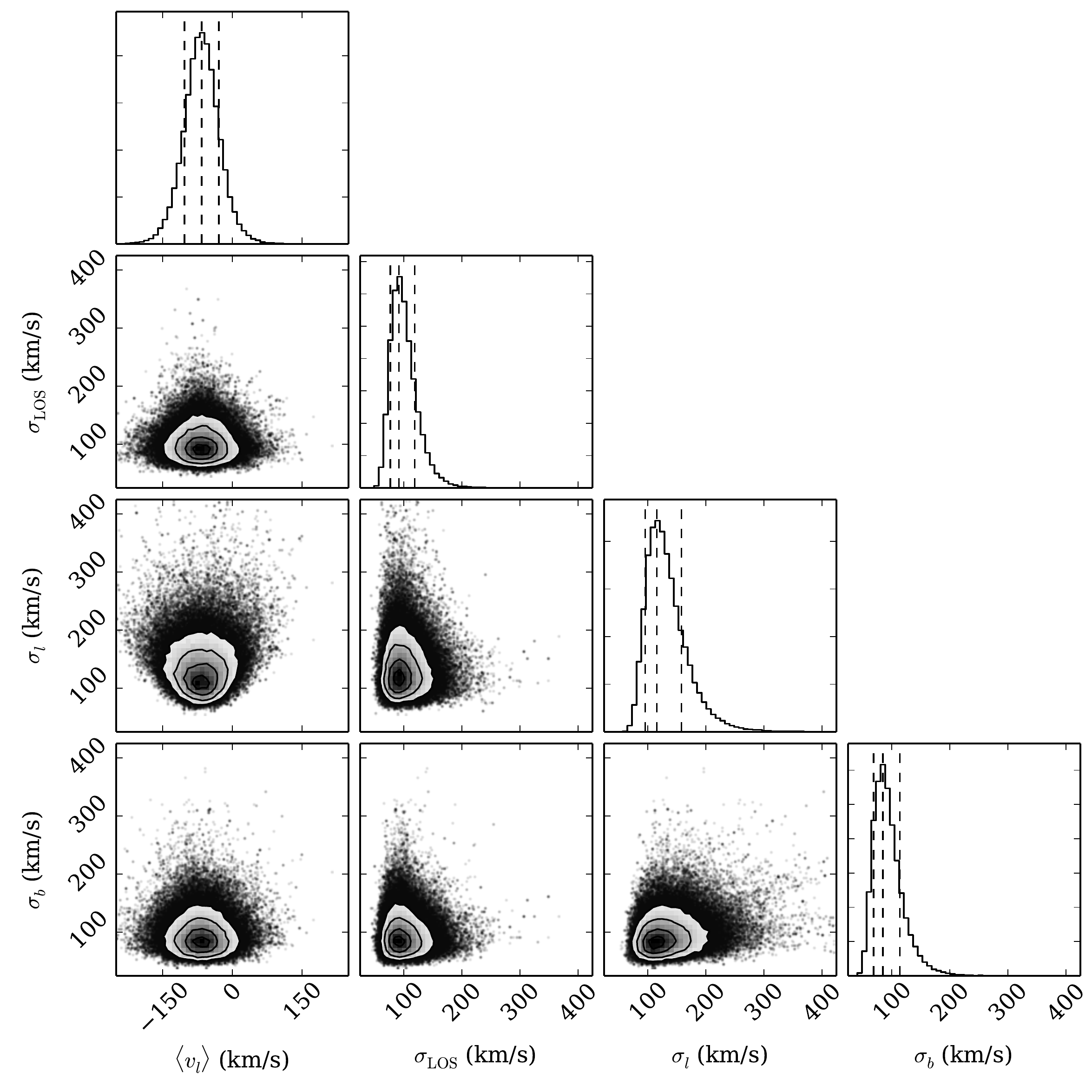}
	\caption[]{Projections of the posterior probability distribution for our four free parameters, when the full sample of 13 objects was used. Contours are shown at 0.5, 1, 1.5 and 2 $\sigma$, respectively. The top panel in each column shows the 1D marginalized PDF for each parameter, with peaks and 68 \% confidence intervals indicated by dashed vertical lines. We acknowledge the use of triangle.py (\citealt{triangle}) to produce this figure.}
	\label{fig:triangle}
\end{figure*}

While the LOS velocity distribution confirms that our sample isn't dominated by TriAnd, TriAnd members could still be biasing our measurement of the anisotropy. Given that our sample is in the same part of the sky and occupies the same region of CMD space as TriAnd (cf. \citealt{Martin2014}), we estimated the TriAnd contamination in our sample by fitting a double Gaussian to the LOS velocity distribution.\footnote{We computed the ratio of evidence (or Bayes factor) to compare the single and double Gaussian models, and found that neither model was strongly favored over the other.} The resulting fit revealed that we expect 2-3 TriAnd stars in our sample, though the underlying hot halo LOS dispersion only changes by $\sim 5 \%$ (see Table \ref{tab:ellipsoid}). The two stars that most likely belong to TriAnd based on their LOS velocities also happen to lie directly over the TriAnd overdensity as seen in CMDs (see Figure 1 of \cite{Martin2014}). The third star with the LOS velocity closest to that of TriAnd lies off the CMD overdensity. We therefore conclude that two of our stars are likely members of TriAnd. The CDF for the double Gaussian best-fit is shown in pink in Figure \ref{fig:vel_hist}.     

Figure \ref{fig:mu_mu} summarizes our 3 dimensional kinematic sample, showing the Galactic proper motion components of the 13 halo stars color coded by LOS velocity. Our sample does not contain any members of M31, as all of these stars have proper motions too large to be associated with M31. As in Figure \ref{fig:vel_hist}, Figure \ref{fig:mu_mu} shows no obvious clumpiness in any kinematic component, indicating that our sample is dominated by a ``hot halo" population. However, it is intriguing that the two stars likely belonging to the TriAnd overdensity (shown as triangles in Figure \ref{fig:mu_mu}) have relatively large proper motions. In the following sections, we consider the halo velocity ellipsoid both with and without the potential TriAnd stars.

\begin{table*}
\begin{center}
\renewcommand{\tabcolsep}{0.2cm}
\renewcommand{\arraystretch}{2}
\begin{tabular}{|c  c  c  c  c|}
\hline
\multicolumn{5}{|c|}{\textbf{Velocity Ellipsoid [km s$^{-1}$]}} \\
Galactic coordinates &&&& \\
\textit{Full Sample} & $\langle v^2_{\rm{LOS}}
\rangle^{1/2}=91^{+27}_{-14}$ &
$\langle v^2_{b} \rangle^{1/2}=88^{+28}_{-17}$ &$\langle
  v^2_{l} \rangle^{1/2}=138^{+43}_{-26}$ & $\langle v_{l}
  \rangle =-67 \pm 37$  \\
\textit{Excluding TriAnd} & $\langle v^2_{\rm{LOS}}
  \rangle^{1/2}=96^{+33}_{-15}$ &
  $\langle v^2_{b} \rangle^{1/2}=82^{+35}_{-16}$ &$\langle
    v^2_{l} \rangle^{1/2}=103^{+50}_{-17}$ & $\langle v_{l}
    \rangle =-50^{+37}_{-40}$  \\
Spherical polar coordinates &&&&\\
\textit{Full Sample} & $\langle v^2_{r}
\rangle^{1/2}=95^{+25}_{-14}$  &
$\langle v^2_{\theta} \rangle^{1/2}=85^{+29}_{-17}$ &$\langle
  v^2_{\phi} \rangle^{1/2}=135^{+41}_{-20}$  & $\langle v_{\phi}
  \rangle = 65 \pm 38 $ \\
\textit{Excluding TriAnd} & $\langle v^2_{r}
  \rangle^{1/2}=100^{+30}_{-15}$  &
  $\langle v^2_{\theta} \rangle^{1/2}=83^{+35}_{-15}$ &$\langle
    v^2_{\phi} \rangle^{1/2}=118^{+50}_{-21}$  & $\langle v_{\phi}
    \rangle = 53 \pm 39 $ \\
\hline
\multicolumn{5}{|c|}{\textbf{Velocity Anisotropy}} \\
\textit{Full Sample} & $\beta=-0.3^{+0.4}_{-0.9}$ & $\sqrt{\frac{ \langle v^2_t
  \rangle}{ \langle v^2_r \rangle}} =
1.6^{+0.5}_{-0.4}$ & $\sqrt{\frac{\langle v^2_\phi \rangle}{ \langle
  v^2_\theta \rangle}}= 1.4^{+0.6}_{-0.4}$ &\\
\textit{Excluding TriAnd} & $\beta=0.1^{+0.4}_{-0.9}$ & $\sqrt{\frac{ \langle v^2_t
    \rangle}{ \langle v^2_r \rangle}} =
  1.4^{+0.6}_{-0.3}$ & $\sqrt{\frac{\langle v^2_\phi \rangle}{ \langle
    v^2_\theta \rangle}}= 1.3^{+0.6}_{-0.3}$ &\\
\hline
\multicolumn{5}{|c|}{\textbf{Position}} \\
& $l=121^\circ$ & $b=-21^\circ$ & $\langle D \rangle = 20 \pm
1 \pm 7$ kpc& $\langle r \rangle = 25 \pm 1 \pm 7$ kpc \\
\hline
\end{tabular}
  \caption{\small Summary of our main results. We give the velocity ellipsoid in
  Galactic and spherical coordinate systems and the
  resulting velocity anisotropy, both for when we include all 13 stars and for when we exclude the 2 stars that are likely TriAnd members. We also give the
  approximate location of our three \textit{HST} fields in the plane
  of the sky, as well as the average heliocentric and
  Galactocentric distances for our sample (which are unchanged to within 0.5 kpc when TriAnd members are excluded). For the latter quantities we list two uncertainties, the first being the error in the mean, and the second being the root-mean-square spread of the sample.}
\label{tab:ellipsoid}
\end{center}
\end{table*}

\section{Velocity Ellipsoid Parameter Estimation}
\label{sec:param}

We use a model of the halo probability distribution function (PDF) to estimate the parameters of the halo velocity ellipsoid ($\langle v_l\rangle$, $\langle v_b\rangle$, $\langle v_{\rm{LOS}}\rangle$, $\sigma_l$, $\sigma_b$, $\sigma_{\rm{LOS}}$) from the observables ($m_{\rm F814W}, m_{\rm F606W}-m_{\rm F814W},\mu_l,\mu_b, l, b, v_{\rm{LOS}}$). The method described is nearly identical to that in D13, though we have made modifications to incorporate the available LOS velocities. We summarize the key points here; see Section 3 of D13 for further details.

First, we determine the PDF for the heliocentric distance to each star. Continuous, double-Gaussian PDFs of absolute magnitude as a function of color were derived using IMF, metallicity, and age weighted \cite{Vandenberg2006} isochrones. We assume a Salpeter IMF, a Gaussian metallicity distribution with mean $[\mathrm{Fe/H}] =-1.9$ and dispersion $\sigma = 0.5$ (e.g., \citealt{Xue2008}), and a Gaussian age distribution with mean $\langle T \rangle = 12$ Gyr and dispersion $\sigma = 2$ Gyr (e.g., \citealt{Kalirai2012}). Possible systematics arising from these assumptions are explored in D13 (see Section 4.2). The resulting absolute magnitude PDF is given by: 

\begin{eqnarray}
\label{eq:abm}
G(M_{\rm F814W} | m_{\rm F606W} - & m_{\rm F814W})=
G_1(A_1,M_1,\sigma_1,M_{\rm F814W}) \nonumber \\
&+ G_2(A_2, M_2, \sigma_2,M_{\rm F814W}),
\end{eqnarray}
where $G(A, M, \sigma,x)=A \, \mathrm{exp}\left[-\left(x-M\right)^2/(2
  \sigma^2)\right]$ and $A$, $M$ and $\sigma$ (amplitude, mean and sigma) are polynomial functions of
  $m_{\rm F606W}-m_{\rm F814W}$ color. See Section 3.1 and Figures 5 and 6 in D13 for more detail. This absolute magnitude PDF is then translated into a distance PDF for each star in our sample using the distance modulus: $D=D(M_{\rm F814W}, m_{\rm F814W})$. 

We then compute the velocity distribution function: $F_v=F_v(v_{\rm{LOS}}, D,\mu_l,\mu_b)$. We assume that the velocity distributions in both tangential and radial directions are Gaussian, with constant values of the ellipsoid parameters over the physical range spanned by our data. We convert observed heliocentric $(v_{l}, v_b)$ velocities to the Galactocentric frame as outlined in Section \ref{sec:vel_meas}. In the direction of
M31, the velocity of the sun projects to: $(v_{l}, v_b) =
(-139.5,83.7)$ ￿. The 3-dimensional velocity probability distribution is given by:

\begin{equation}
\begin{split}
	F_v(v_l, v_b, v_{\rm{LOS}}) = \frac{1}{\left(2 \pi\right)^{3/2} \sigma_l \sigma_b \sigma_{\rm{LOS}}}  \mathrm{exp}\left[-\frac{\left(v_l-\langle v_l \rangle \right)^2}{2
	    \sigma^2_{l}}\right] \\ 
		\times \mathrm{exp}\left[-\frac{\left(v_b-\langle v_b \rangle \right)^2}{2
	    \sigma^2_{b}}\right] \mathrm{exp}\left[-\frac{\left(v_{\rm{LOS}}-\langle v_{\rm{LOS}}\rangle \right)^2}{2
	    \sigma^2_{\rm{LOS}}}\right].
\end{split}
\end{equation}

The halo PDF at fixed $m_{\rm F606W}-m_{\rm F814W}$ color, in increments of absolute magnitude, apparent magnitude, Galactic PM, LOS velocity and solid angle ($\Omega$), $F(y)$, where $y$ is defined as $y=y(M_{\rm F814W}, m_{\rm F814W}, \mu_l, \mu_b, v_{\rm{LOS}}, \Omega$), is given by:

\begin{equation}
\label{eq:full_pdf}
F \, \Delta \textbf{y}  = F_v \, \rho \, D^5 \, G \, \,\cos(b) \Delta \textbf{y}. 
\end{equation}

Here, $\rho=\rho(D, l,b)$ is the density distribution of halo stars (we assume the broken power law profile derived by \cite{Deason2011b}), $G=G(M_{\rm F814W} | m_{\rm F606W} -
m_{\rm F814W})$ is the absolute magnitude PDF in Eqn. \ref{eq:abm} and $\Delta
\textbf{y}= \Delta M_{\rm F814W} \Delta m_{\rm F814W} \Delta \mu_l \Delta \mu_b \Delta v_{\rm{LOS}} \Delta \Omega$ is the volume element. 

We marginalize over absolute magnitude, and define the likelihood function:

\begin{equation}
\label{eq:like}
L=\prod \bar{F}(\sigma_l, \sigma_b, \sigma_{\rm{LOS}}, v_{l,0}, v_{b,0}, v_{LOS,0}, \textbf{x}),
\end{equation}
where $\bar{F} = \int F \, \mathrm{d} M_{\rm F814W}$. 

We sample the marginalized posterior probability distribution with \verb+emcee+ (\citealt{ForemanMackey2013}), a \textsc{python} implementation of the \cite{Goodman2010} affine-invariant Markov chain Monte Carlo (MCMC) ensemble sampler. We set $\langle v_b \rangle=\langle v_{\rm{LOS}} \rangle =0$, but allow for net motion in Galactic longitude, which approximates the net rotational velocity ($v_{\phi}$) of the halo. We assume a flat prior on the mean velocity in galactic longitude $\langle v_{l} \rangle$ and a flat prior between 0 and 450 $\rm{km~s^{-1}}$ on the dispersions. Projections of our posterior probability are shown in Figure \ref{fig:triangle}.

\section{Results}
\label{sec:res}

Figure \ref{fig:triangle} shows projections of the samples of the posterior, with marginalized one-dimensional PDFs for each parameter shown in the top panel of each column. We find the following values for the velocity ellipsoid parameters, with $68\%$ confidence limits:
$\langle v_l \rangle=-66^{+37}_{-37} \ \mbox{km \ s}^{-1}$, $\sigma_{\rm{LOS}}=91^{+27}_{-14} \ \mbox{km \ s}^{-1}$, $\sigma_{l}=117^{+42}_{-19} \ \mbox{km \ s}^{-1}$, and $\sigma_{b}=88^{+28}_{-17} \ \mbox{km \ s}^{-1}$. Here we have quoted the peaks of the 1D marginalized PDFs, and the limits enclose 68\% of the points on either side of the peak.

We convert our velocity ellipsoid quantities to spherical polar coordinates using a Monte Carlo method. Our galactocentric polar coordinate system is defined such that the sun is located on the negative $x$ axis, and the polar angle $\phi$ is the angle from the negative $x$ axis to the positive $y$ axis ($l=90 \degree$), such that $\phi$ is positive in the direction of Galactic rotation. To make the conversion from $v_l,\ v_b,\ v_{\rm{LOS}}$ to $v_r, \ v_{\theta}, \ v_{\phi}$, we generate a random sample of $\sim 25,000$ stars drawn from the halo density distribution (\citealt{Deason2011b}):

\begin{equation}
	\rho(r_q) \propto \left\{ 
	\begin{array}{lr}
		r_q^{-\alpha_{\rm{in}}} & r_q \leq r_b,\\
		r_q^{-\alpha_{\rm{out}}} & r_q > r_b. \\
	\end{array}
	\right.
\end{equation}
where $r_q=x^2+y^2+z^2q^{-2}$, $q=0.59$ is the halo flattening parameter, $r_b=27$ kpc, $\alpha_{\rm{in}}=2.3$, and $\alpha_{\rm{out}}=4.6$. The stars are placed along the line-of-sight and have heliocentric distances ranging from 10 to 100 kpc. The stars are assigned a velocity distribution based on a random selection from our MCMC samples. Each star's velocity components $v_r, v_{\theta}, v_{\phi}$ are calculated from the generated positions and $v_l$, $v_b$, $v_{\rm{LOS}}$ velocities. The second moments in spherical polar coordinates are computed from the resulting Galactocentric velocity distributions.

By repeating this process $10^5$ times, we compute PDFs for the second moments for the galactocentric velocity ellipsoid parameters. The uncertainties on these parameters are computed in the same way as the heliocentric velocity ellipsoid parameters: the limits enclose 68\% of the points on either side of the peak. Our results are summarized in Table \ref{tab:ellipsoid}. Using the PDFs for the galactocentric second moments, we compute the PDF for the anisotropy parameter (\citealt{Binney2008}):

\begin{equation}
	\beta=1-\frac{\langle v_{\theta}^{2}\rangle+\langle v_{\phi}^{2}\rangle}{2\langle v_r^{2}\rangle}.
\end{equation}

We find $\beta=-0.3^{+0.4}_{-0.9}$, where we again quote the peak of the PDF and limits that enclose 68\% of the points on either side of the peak. If we repeat this analysis excluding the two likely TriAnd members, we find $\beta=0.1^{+0.4}_{-0.9}$. Both of these values are consistent with the value found in D13 ($\beta_{D13} = 0.0 ^{+0.2}_{-0.4}$), though our new values have larger error bars because we measured the LOS velocity distribution directly. Our values for the ellipsoid parameters in this case are also quoted in Table \ref{tab:ellipsoid}. 

\begin{figure*}
	\centering
	\includegraphics[width=\textwidth]{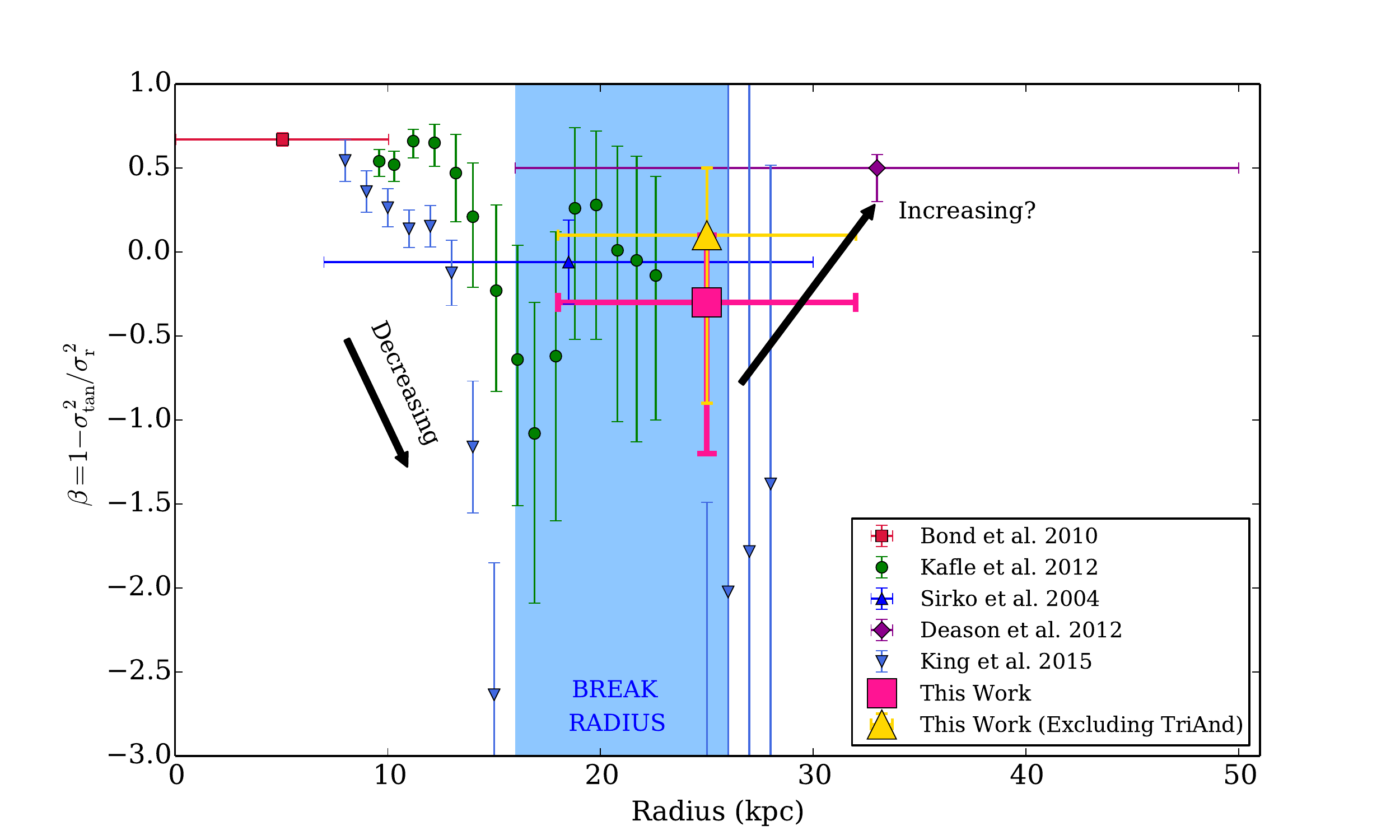}
	\caption[]{Radial velocity anisotropy profile. The ``break radius" of the Milky Way stellar halo is shown by the blue shaded region ($16 \lesssim r/\rm{kpc} \lesssim 26$; \citealt{Deason2011b}). Our measurement of $\beta$, from the 3D kinematics of $N=13$ stars in the radial range $18 \lesssim r/\rm{kpc} \lesssim 32$, is shown in pink. Solar neighborhood measurements, using full 3D velocity information, find a radially biased $\beta$, shown in red (\citealt{Bond2010}; \citealt{Smith2009}). The remaining error bars show estimates of $\beta$ for distant ($D \gtrsim 10$ kpc) halo stars using LOS velocity distributions (\citealt{Sirko2004}; \citealt{Kafle2012}; \citealt{Deason2012}; \citealt{King2015}). }
	\label{fig:beta}
\end{figure*}

\section{Discussion}
\label{sec:disc}

Our value of $\beta$ is consistent with isotropy, and lower than local measurements by at least $1\sigma$, which find a radially biased anisotropy ($\beta = 0.5-0.7$). The significant dispersion in the observed LOS velocity distribution (Figure \ref{fig:vel_hist}) rules out the possibility that our sample is dominated by cold substructure. 

Figure \ref{fig:beta} shows the radial anisotropy profile of the Milky Way stellar halo. Our measurement is consistent with the observed ``dip'' in the anisotropy profile, seen in multiple studies that measured the velocity anisotropy of distant halo stars along different lines of sight using only LOS velocity distributions (\citealt{Sirko2004}; \citealt{Kafle2012}; \citealt{Deason2012}; \citealt{King2015}). This dip is also coincident with the observed break in the halo density profile around $16~\mbox{kpc} \lesssim r \lesssim 26~\mbox{kpc}$ (\citealt{Deason2011b}; \citealt{Sesar2011}; \citealt{Watkins2009}). In this section, we discuss some of the possible explanations of this result. 

\subsection{A Galactic Shell}

In D13, we argued that the presence of global substructure, such as a shell (or multiple shells), is one explanation for both the steep fall-off in stellar density beyond the break radius and the decrease in anisotropy at that radius. \cite{Deason2013a} argued that a break in the Milky Way stellar density profile could be created by the build-up of stars  at apocenter from either one relatively massive accretion event or several, synchronous accretion events. In this scenario, we would expect the stars to have an increase in tangential motion relative to radial motion at the turnaround radius, and thus a more isotropic $\beta$, just as we observe. This picture is consistent with what we find for likely TriAnd members: TriAnd is a large, cloud-like overdensity of stars likely at apocenter (\citealt{Johnston2012}), and including TriAnd in our sample makes $\beta$ more tangentially biased. Chemical abundances for these stars may help to characterize the progenitor (or progenitors) of this shell (see Section \ref{sec:chem}).

Several of these cloud-like overdensities, such as TriAnd, the Virgo overdensity (VOD), the Hercules-Aquila overdensity (HerAq) and the Eridanus-Phoenix overdensity (EriPhe) are all located at approximately 20 kpc. \cite{Li2015} recently suggested that EriPhe, HerAq and the VOD could all be associated, and potentially fell in to the Milky Way as a group; TriAnd could also be a member of this group. A group infall event could explain the presence of all these overdensities at $\sim$ 20 kpc, the observed break in the density profile and the relative increase in tangential motion at this radius.  

\subsection{Dual Stellar Halo: In-Situ Star Formation}

The break in the stellar density profile could also be an indication that the Milky Way has a ``dual stellar halo,'' containing populations of different origins (\citealt{Carollo2007}; \citealt{Carollo2010}; \citealt{Beers2012}). Simulations predict that the stellar halo is composed both of accreted stars and stars that form \textit{in-situ} (e.g., \citealt{Zolotov2009}; \citealt{Font2011}; \citealt{McCarthy2012}; \citealt{Tissera2012}; \citealt{Cooper2015}). In-situ stars have two flavors. The first are stars that form in the halo itself from gas accreted from the IGM or satellites; it remains unknown to what extent these populations and their properties are a result of the choice of hydrodynamics scheme. Secondly, stars can form in the disk of the Milky Way and then be kicked up into the halo due to merger events (these stars are sometimes called ``heated disk stars''). In simulations, these stars can comprise a significant fraction of the stellar population (and sometimes even dominate) within $r \lesssim 30$ kpc. It's possible that our observed isotropy is a kinematic signature of a heated disk population. \cite{McCarthy2012} showed that these
in-situ stars can have significant prograde rotation and
therefore increased tangential pressure support from angular
momentum, and we find a significant signal of prograde rotation ($\langle v_{\phi}\rangle \sim 70 \ \mbox{km \ s}^{-1}$). However, this scenario does not explain why measurements of the velocity anisotropy in the solar neighborhood find radially biased orbits, in the region of the halo where we would expect even more heated disk stars. Distinguishing between accreted and in-situ populations with kinematics alone remains challenging, and model predictions remain unclear. To better determine if our objects were accreted or formed in-situ, we need chemical abundances (see Section \ref{sec:chem}).

\subsection{Future Work}

\subsubsection{Chemistry}
\label{sec:chem}

Chemical information is key for disentangling the Milky Way's accretion history. Iron abundances of accreted populations are related to the masses of the dwarf progenitors (e.g., see \citealt{Johnston2008}; \citealt{Kirby2013}). If our 13 stars are accreted halo stars, measuring iron abundances may help to determine whether a single accretion event or several are responsible for the shell-type structure we observe, and we can use the abundances to estimate the mass(es) of the progenitor(s). 

The chemical information in our stellar spectra is also our best hope of determining the relative contributions of different stellar halo formation mechanisms. Stars that form in the disk of the Milky Way in simulations are found to have a higher average [Fe/H] than accreted stars (\citealt{Font2011}; \citealt{Tissera2012}; \citealt{Cooper2015}). In addition, \cite{Zolotov2010} showed that in-situ stars are  alpha-enriched relative to accreted stars at a given [Fe/H] at the high [Fe/H] end of the metallicity distribution function. These results are due to the fact that in-situ stars form in a deeper potential well than the accreted population. Several studies have used abundances in an effort to disentangle these populations locally (e.g. \cite{Nissen2010}, with F and G main sequence stars within 335 pc; \cite{Sheffield2012} with M Giants out to 10 kpc). However, no such studies exist using main sequence stars outside the solar neighborhood. By measuring the iron and alpha abundances of distant main sequence halo stars, we can begin to assess the relative importance of different physical processes leading to the formation of the Milky Way's stellar halo.  

\subsubsection{HALO7D}
In order to better understand the global halo properties, we need more than $N\sim13$ stars! Through the HALO7D observing program (begun in Spring 2014), we will obtain deep (8--24 hour integrations) spectra of hundreds of distant MSTO halo stars \textit{with measured HST proper motions} using Keck II/DEIMOS. We will target $N\sim350$ stars in the four CANDELS fields (\citealt{Grogin2011}; \citealt{Koekemoer2011}): GOODS-N, GOODS-S, COSMOS, and EGS. All four of these fields are characterized by deep, multi-epoch \textit{HST} imaging, and cover a total area of approximately 1000 square arcminutes. With this dataset, we will:

\begin{enumerate}
	\item Measure LOS velocities of all stars, as well as [Fe/H] and [$\alpha$/Fe] for those stars with sufficient signal to noise.
	\item Measure the velocity anisotropy along four new lines of sight.
	\item Measure the anisotropy as a function of galactocentric distance exclusively with stars that have 3D kinematic information. 
	\item Use chemical abundances to disentangle the Milky Way's accretion history and determine the relative contributions of stellar halo formation mechanisms.
\end{enumerate}

HALO7D is an ongoing observational program with results forthcoming (Cunningham et al., in prep). 

\section{Conclusions}
\label{sec:concl}

We present line-of-sight (LOS) velocities for $N=13$ Milky Way halo stars with measured \textit{HST} proper motions (PMs). Our sample is the first sample of halo stars with measured 3D kinematics outside of the solar neighborhood. The LOS velocities were measured from deep (5-6 hour) integrations on Keck II/DEIMOS. We combine the LOS velocity measurements with the proper motions to estimate the parameters of the velocity ellipsoid using an MCMC ensemble sampler. We find the velocity distribution in Galactic longitude \textit{l} to have a mean $\langle v_{l} \rangle = -67^{+37}_{-37}\ \mbox{km \ s}^{-1}$ and a dispersion $\sigma_{l}= 117^{+42}_{-20} \ \mbox{km \ s}^{-1}$. We find the dispersions in Galactic latitude \textit{b} and the LOS to be $\sigma_{b}=88^{+28}_{-17} \ \mbox{km \ s}^{-1}$ and $\sigma_{\rm{LOS}}= 91^{+27}_{-14} \ \mbox{km \ s}^{-1}$, respectively. 

Using our estimates of the ellipsoid parameters, we measure the velocity anisotropy $\beta$. We find $\beta=-0.3^{+0.4}_{-0.9}$, consistent with isotropy and with the result from D13, but lower than solar neighborhood measurements, which find a radially biased $\beta$, by at least 1$\sigma$. If we exclude likely TriAnd members from our sample, we find $\beta=0.1^{+0.4}_{-0.9}$. These values are also consistent with other observational studies (using only LOS velocities) that have found a decrease in $\beta$ around the observed break radius in the Milky Way density profile ($16~\mbox{kpc} \lesssim r \lesssim 26~\mbox{kpc}$). These two findings in tandem suggest the presence of a shell-type structure in the halo at this radius, potentially formed by several destroyed dwarfs with similar apocenters. It is also possible that we are observing a population dominated by \textit{in-situ} stars rather than an accreted population. 

We need more observations and chemical information for distant halo stars to better understand the origin of the Milky Way stellar halo and its accretion history. We will achieve this with the HALO7D observing program, which will increase our sample of stars with 3D kinematics by a factor of $\sim$ 30. The velocities and abundances measured from these observations will vastly improve our understanding of the Galaxy's accretion history and the origin of the stellar halo.

\section*{Acknowledgments}
We thank the anonymous referee for the helpful comments and suggestions. ECC is supported by a National Science Foundation Graduate Research Fellowship. AJD is supported by a Porat Fellowship at Stanford University. ET acknowledges the National Science Foundation grants AST-1010039 and AST-1412504. Partial support for this work was provided by NASA through grants for program AR-13272 from the Space Telescope Science Institute (STScI), which is operated by the Association of Universities for Research in Astronomy (AURA), Inc., under NASA contract NAS5-26555. ECC thanks Charles King III and Warren Brown for kindly providing their data for Figure \ref{fig:beta}. ECC thanks Alexa Villaume, Christopher Mankovich and Zachary Jennings for helpful scientific conversations. We thank the outstanding team at Keck Observatory for assisting us in our observations. This research made use of Astropy, a community-developed core Python package for Astronomy (Astropy Collaboration, 2013). We recognize and acknowledge the significant cultural role and reverence that the summit of Mauna Kea has always had within the indigenous Hawaiian community. We are most fortunate to have the opportunity to conduct observations from this mountain. 

\vspace{0.5cm}

\label{lastpage}
\bibliography{m31bib}

\end{document}